\documentclass[reprint,twocolumn,aps,pra,notitlepage,showpacs,floatfix,amsmath]{revtex4-1}
\usepackage{graphicx}
\usepackage{amssymb}
\usepackage{epstopdf}
\usepackage{amsmath,amsfonts,latexsym}
\newcommand{\vect}[1]{\mathbf{#1}}

\def\pra{\ref@jnl{Phys.~Rev.~A}}
\def\prb{\ref@jnl{Phys.~Rev.~B}}
\def\prc{\ref@jnl{Phys.~Rev.~C}}
\def\prd{\ref@jnl{Phys.~Rev.~D}}
\def\pre{\ref@jnl{Phys.~Rev.~E}}
\def\prl{\ref@jnl{Phys.~Rev.~Lett.}}

\usepackage{amsmath,amsthm,amsfonts,amssymb,mathrsfs,mathtools,bbm,cancel}
\usepackage{enumerate}
\usepackage{color}
\usepackage{pst-func}
\usepackage{dcolumn}
\usepackage{bm}
\usepackage[bookmarks=false,hyperfootnotes=false]{hyperref}  
\hypersetup{
			colorlinks=true,
			linkcolor=red,
			urlcolor=blue,
			anchorcolor=black,
			citecolor=blue,
			urlcolor=black,
			pdftitle={Breathing mode frequency of a strongly %
				interacting Fermi gas across %
				the 2D-3D dimensional crossoverr},
			pdfauthor={Umberto Toniolo}
}

\DeclareMathOperator{\arcsinh}{arcsinh}
\DeclareMathOperator{\im}{i \hspace{-0.2mm}}

\newcommand{\ZZ}{\mathbb{Z}}
\newcommand{\RR}{\mathbb{R}}
\renewcommand{\epsilon}{\varepsilon}
\newcommand{\kft}{k_F^{\text{3D}}}
\newcommand{\kfw}{k_F^{\text{2D}}}

\newcommand{\eft}{\epsilon_F^{\text{3D}}}

\newcommand{\ef}{\epsilon_F}
\newcommand{\bzero}{B_{0}}
\newcommand{\lz}{l_z}
\newcommand{\peta}{\kft\lz}
\newcommand{\at}{a_{\text{3D}}}
\newcommand{\aw}{a_{\text{2D}}}
\newcommand{\lza}{\lz/\at}

\begin{document}

\title{Breathing mode frequency of a strongly interacting Fermi gas across
the 2D-3D dimensional crossover}

\author{Umberto Toniolo}

\author{Brendan C. Mulkerin}

\author{Xia-Ji Liu}

\author{Hui Hu}

\affiliation{Centre for Quantum and Optical Science, Swinburne University of Technology,
Hawthorn 3122 VIC, Australia}

\date{\today}
\begin{abstract}
We address the interplay between dimension and quantum anomaly on
the breathing mode frequency of a strongly interacting Fermi gas harmonically
trapped at zero temperature. Using a beyond mean-field, Gaussian pair
fluctuation theory, we employ periodic boundary conditions to simulate
the dimensionality of the system and impose a local density approximation,
with two different schemes, to model different trapping potentials
in the tightly-confined axial direction. By using a sum-rule approach,
we compute the breathing mode frequency associated with a small variation
of the trapping frequency along the weakly-confined transverse direction,
and describe its behavior as functions of the dimensionality, from
two- to three-dimensions, and of the interaction strength. We compare
our predictions with  previous calculations on the two-dimensional
breathing mode anomaly and discuss their possible observation in ultracold
Fermi gases of $^{6}$Li and $^{40}$K atoms.
\end{abstract}
\maketitle

\section{Introduction}

Low-lying collective excitations play a fundamental role in understanding
many-body phenomena and the recent realization of ultracold atomic
gases provides a unique setting for investigating various novel collective
dynamics \cite{Bloch2008}. In particular, low dimensional atomic
Fermi gases (in one and two dimensions) at the crossover from a Bose-Einstein
condensate (BEC) to a Bardeen-Cooper-Schrieffer (BCS) superfluid present
a broad range of intriguing collective phenomena that are now being
successfully studied from both theoretical and experimental perspectives
\cite{Vogt2012,Hofmann2012,Gao2012,Taylor2012,Levinsen2015,Mulkerin2017}.
Low-dimensional regimes are experimentally achieved by using a combination
of harmonic oscillator (HO) traps \cite{Bloch2008,Chin2010}, whose
oscillating frequencies are tuned in order to reach both two-dimensional
(2D) pancake \cite{Fenech2016,Hueck2018} and one-dimensional (1D)
cigar traps \cite{Guan2013}. Alternatively, a standing wave laser
beam in a given selected direction can force the system into a quasi-2D
pancake-like regime, where multiple almost independent 2D clouds are
realized \cite{Ong2015,Cheng2016,Boettcher2016,Turlapov2017}. 

It is well known that the long-range order parameter is expected to
be highly suppressed by fluctuations in low dimensions, however it
is still possible to achieve superfluidity with a quasi-long-range
order according to the Berezinskii-Kosterlitz-Thouless phase transition
universality \cite{Berezinskii,Kosterlitz1972}. The interest around
low-dimensional quantum gases, in particular for the 2D case, is further
emphasized by features that are known to have no classical counterpart.
At low temperatures an interacting Fermi gas experiences mainly $s$-wave
scattering, which are theoretically modeled by a contact interaction.
It is straightforward to observe that in two dimensions the Hamiltonian
of such a system is invariant under length scaling, however due to
the contact interaction unphysical contributions at large momenta 
are included. The solution to this problem is to introduce an extra
length scale upon renormalization, the scattering length $a_{{\rm 2D}}$,
which breaks the scale invariance of the classical Hamiltonian and
leads to the phenomenon of the so-called \emph{quantum anomaly} \cite{Holstein1993}.

A well-known consequence due to the breaking of scale invariance in
two-dimensional gases can be found when exciting the harmonically
trapped cloud via collective excitations. Namely, a small perturbation
of the transverse harmonic frequency $\omega_{\perp}$ induces a breathing
mode excitation whose frequency is given by $\omega_{B}=2\omega_{\perp}$
\cite{Rosch97,Werner2006}. This classical result is modified when
the quantum anomaly is considered. The breathing mode frequency gains
a \emph{weak} dependence on the scattering length and deviates from
the classical value, $\omega_{B}=2\omega_{\perp}$, within a range
of $5-10\%$ \cite{Olshanii2010,Hofmann2012,Gao2012}. This should
be contrasted with the case of a three-dimensional gas, in which the
classical Hamiltonian is in general not invariant under length scaling.
Due to the contact interaction we must also renormalize the Hamiltonian
with the 3D scattering length, $\at$, and the breathing mode thus
\emph{strongly} depends on the scattering length. The only exception
is the unitarity limit, where the scattering length diverges, $\at=\pm\infty$,
and the quantum Hamiltonian becomes scale invariant. As a result of
the restored scale invariance, the breathing mode does not depend
on temperature \cite{Hou2013}. In the 3D regime, for a unitary Fermi
cloud in the highly pancake-like trapping potential, the scale invariant
breathing mode takes $\omega_{B}=\sqrt{3}\omega_{\perp}$ \cite{Hu2014,DeRosi2015}.
It is of great interest to study how the breathing mode frequency
evolves at the dimensional crossover from 3D to 2D, while aiming for
the realization of a truly 2D gas.

In this work, motivated by the recent experimental %
activities at Swinburne University of Technology~\cite{Peppler2017}, %
we address the role of dimension and interaction on
the breathing mode frequency and the quantum anomaly. The suppressed
superfluid order parameter in low dimensions requires a beyond mean-field
(MF) treatment \cite{tempere2012,Mulkerin2017a}, which is possible
when we consider periodic boundary conditions (PBC) in the axial direction
\cite{Toniolo2017b}. Moreover, the effect of harmonic trapping in
the transverse direction on the integrated 2D density distribution
can be well described by a local density approximation (LDA). We describe
the breathing mode frequency for a given pair of parameters: a length
that tunes the dimension and a scattering length that sets the interaction
strength. By using a sum-rule approach, in the spirit of Ref. \cite{Menotti2002},
we determine the breathing mode frequency while changing the dimensional
regime and tuning the scattering length. We further address a comparison
with the previous results of the breathing mode frequency in the purely
2D regime \cite{Hofmann2012,Gao2012,Taylor2012}. Our predictions
could be readily examined in future cold-atom experiments with fermionic
$^{6}$Li and $^{40}$K atoms.

The paper is set out as follows: in Sec. \ref{ssec:dimensionalcrossover}
we go through the beyond-MF, Gaussian pair fluctuation (GPF) theory
to study a homogeneous strongly interacting Fermi gas with periodic
boundary conditions, and in Sec. \ref{ssec:lda} we introduce two
different LDA schemes to account for different axial confinements.
In Sec. \ref{ssec:breathingmode} we briefly derive the sum-rule for
the breathing mode frequency calculations, which are extensively addressed
in Appendix \ref{app:schemes}, for the sake of clarity. Finally in
Sec. \ref{sec:results} we show the behavior of the breathing mode
frequency in various dimensional regimes and the comparison with previous
2D results.

\section{Theoretical models}

\label{sec:model} 

\subsection{Homogeneous strongly interacting Fermi gases at the dimensional crossover}

\label{ssec:dimensionalcrossover} Following our previous work in
Ref.~\cite{Toniolo2017b}, we model a two-component spin balanced
Fermi gas near a broad Feshbach resonance via a two-body contact interaction.
In order to describe the dimensional crossover, we split the spatial
coordinates into in-plane $\vect{x}=(x,y)$ and axial $z$ components,
combined in the short-hand notation $(\vect{x},z)$. The grand canonical
single channel Hamiltonian reads \cite{Randeria1989}, 
\begin{equation}
\begin{split}\mathcal{H}=\sum_{\sigma=\uparrow,\downarrow} & \psi_{\sigma}^{\dagger}(\vect{x},z)\left(-\frac{\hbar^{2}}{2m}\nabla^{2}-\mu\right)\psi_{\sigma}(\vect{x},z)\\
 & -U_{0}\psi_{\uparrow}^{\dagger}(\vect{x},z)\psi_{\downarrow}^{\dagger}(\vect{x},z)\psi_{\downarrow}(\vect{x},z)\psi_{\uparrow}(\vect{x},z),
\end{split}
\label{eq:hamiltonian}
\end{equation}
where $\psi_{\sigma}$ are the annihilation field operators for the
(pseudo-)spin populations labeled by $\sigma=\uparrow,\downarrow$,
$\mu$ is the chemical potential, $m$ is the mass of the fermions,
and $U_{0}>0$ is the bare interaction strength of the contact potential.
We introduce the Hubbard-Stratonovich auxiliary bosonic field, $\hat{\Delta}(x)=U_{0}\psi_{\downarrow}(\vect{x},z)\psi_{\uparrow}(\vect{x},z)$,
and we perform a saddle point approximation with the mean-field order
parameter and the fluctuation bosonic field \cite{Hu2006,Diener2008},
$\hat{\Delta}(x)=\Delta+\hat{\phi}(\vect{x},z)$. This approximation
allows us to directly compute the thermodynamic potential up to second
order in the fluctuation, $\Omega=\Omega_{{\rm MF}}+\Omega_{{\rm GPF}}$,
where at the MF level we have, 
\begin{equation}
\frac{\Omega_{\text{MF}}}{V}=\frac{\Delta^{2}}{U_{0}}+\frac{1}{V}\sum_{\vect{k},k_{z}}(\xi_{\vect{k},k_{z}}-E_{\vect{k},k_{z}}),
\end{equation}
where $V$ is the volume of the system and we have introduced the
BCS theory notation, slightly modified for the dimensional crossover,
for a generic momentum $(\vect{k},k_{z})$: we define $\xi_{\vect{k},k_{z}}=\epsilon_{\vect{k}}+\epsilon_{k_{z}}-\mu$
and $E_{\vect{k},k_{z}}=\sqrt{\xi_{\vect{k},k_{z}}^{2}+\Delta^{2}}$.
The GPF contribution to the thermodynamic potential, at finite temperature
$k_{B}T=\beta^{-1}$, is 
\begin{equation}
\Omega_{\text{GPF}}=-\frac{1}{\beta}\ln\int\mathcal{D}\phi^{*}\mathcal{D}\phi\exp\left[\mathcal{S}_{\text{GPF}}(\phi^{*},\phi)\right],\label{eq:path}
\end{equation}
where $k_{B}$ is the Boltzmann constant and the GPF action, $\mathcal{S}_{\text{GPF}}$,
can be written as 
\begin{equation}
\mathcal{S}_{\text{GPF}}=\frac{\beta V}{2}\sum_{Q}\begin{pmatrix}\phi_{Q}^{*} & \phi_{-Q}\end{pmatrix}\mathbf{M}(Q)\begin{pmatrix}\phi_{Q}\\
\phi_{-Q}^{*}
\end{pmatrix}.
\end{equation}
We have introduced the multi-index notation $Q\equiv(\vect{q},q_{z},\im q_{\nu})$
with momenta, $(\vect{q},q_{z})$, of the fluctuation field, $\phi$,
and the bosonic Matsubara frequencies $q_{\nu}=2\pi\nu/\beta$, for
all $\nu\in\ZZ$. The matrix operator $\mathbf{M}$ at zero temperature
can be written, 
\begin{eqnarray}
\mathbf{M}_{11} & = & \frac{1}{U_{0}}+\frac{1}{V}\sum_{\mathbf{k},k_{z}}\left(\frac{u_{+}^{2}u_{-}^{2}}{\im q_{\nu}-E_{+}-E_{-}}-\frac{v_{+}^{2}v_{-}^{2}}{\im q_{\nu}+E_{+}+E_{-}}\right),\nonumber \\
\mathbf{M}_{12} & = & \frac{1}{V}\sum_{\mathbf{k},k_{z}}\left(-\frac{u_{+}u_{-}v_{+}v_{-}}{\im q_{\nu}-E_{+}-E_{-}}+\frac{u_{+}u_{-}v_{+}v_{-}}{\im q_{\nu}+E_{+}+E_{-}}\right),
\end{eqnarray}
$\mathbf{M}_{21}\left(Q\right)=\mathbf{M}_{12}\left(Q\right)$ and
$\mathbf{M}_{22}\left(Q\right)=\mathbf{M}_{11}\left(-Q\right)$. Here,
we use the notations \cite{Hu2006,Diener2008,He2015} 
\begin{equation}
u_{\pm}^{2}=1-v_{\pm}^{2}=\frac{1}{2}\left(1+\frac{\xi_{\pm}}{E_{\pm}}\right),
\end{equation}
with $E_{\pm}=\sqrt{\xi_{\pm}^{2}+\Delta^{2}}$ and $\xi_{\pm}\equiv\xi_{\mathbf{k}\pm\frac{\mathbf{q}}{2},k_{z}\pm\frac{q_{z}}{2}}$.
At zero temperature, for $\beta\rightarrow\infty$, we can Wick rotate
the Matsubara frequencies \cite{Diener2008,He2015}, $\im q_{\nu}\mapsto\omega$,
swapping the sum on $\im q_{\nu}$ with an integration on $\omega$,
\begin{equation}
\Omega_{\text{GPF}}=\frac{1}{V}\sum_{\vect{q},q_{z}}\int_{0}^{\infty}\frac{d\omega}{2\pi}\ln\Gamma^{-1}(\vect{q},q_{z},\omega),
\end{equation}
where 
\begin{equation}
\Gamma^{-1}(Q)=\frac{\mathbf{M}_{11}(Q)\mathbf{M}_{11}(-Q)-\mathbf{M}_{12}(Q)^{2}}{\mathbf{M}_{11}^{C}(Q)\mathbf{M}_{11}^{C}(-Q)},\label{eq:vertex}
\end{equation}
and we have introduced an additional term to converge the integrations
\cite{Diener2008,He2015}, 
\begin{equation}
\mathbf{M}_{11}^{C}\left(Q\right)=\frac{1}{U_{0}}+\frac{1}{V}\sum_{\mathbf{k},k_{z}}\frac{u_{+}^{2}u_{-}^{2}}{\im q_{\nu}-E_{+}-E_{-}}.
\end{equation}
The dispersion relation of the bosonic field should be gapless, hence
we determine the order parameter, $\Delta$, at the MF level by solving
the gap equation
\begin{equation}
\mathbf{M}_{11}(Q=0)-\mathbf{M}_{12}(Q=0)=0,
\end{equation}
or more explicitly,
\begin{equation}
\frac{1}{U_{0}}-\frac{1}{V}\sum_{\vect{k},k_{z}}\frac{1}{2E_{\vect{k},k_{z}}}=0.
\end{equation}

In the spirit of Ref. \cite{Toniolo2017b}, we introduce the tuning
parameters of the dimensional crossover as follows: the in-plane coordinates
are sent to the thermodynamic limit, while we require PBC to hold
on the axial direction. The characteristic PBC length $\lz$ tunes
the dimensional crossover from the 3D (large $\lz$) limit towards
the 2D (small $\lz$) regime. We define the characteristic Fermi momentum
$k_{F}$ and Fermi energy $\varepsilon_{F}=\hbar^{2}k_{F}^{2}/(2m)$
from the free Fermi density $n_{f}$. That is, for a fixed box length
$l_{z}$, we take the discretization of momenta in the $z$ direction,
$k_{z}=2\pi N_{z}/l_{z}$, for any integer $N_{z}$. The free Fermi
density is then given by \cite{Toniolo2017b}, 
\begin{equation}
n_{f}=\frac{1}{2\pi l_{z}}\sum_{N_{z}=-N_{\text{max}}}^{N_{\text{max}}}\left[k_{F}^{2}-\left(\frac{2\pi N_{z}}{l_{z}}\right)^{2}\right],\label{eq:densityq2d}
\end{equation}
where $N_{\text{max}}$ is the largest natural number smaller than
$k_{F}l_{z}/(2\pi)$. In the 2D and 3D limits, the Fermi momentum
$k_{F}$ should approach respectively their limiting values, $\kfw$
and $\kft$, which can be defined by the 2D and 3D free Fermi densities
$n_{\text{2D}}=n_{f}l_{z}$ and $n_{\text{3D}}=n_{f}$ in the usual
way. For convenience, we introduce the \emph{dimensional crossover
tuning parameter} via the 3D Fermi momentum $\kft$ \cite{Toniolo2017b}:
\begin{equation}
\eta\equiv\peta.
\end{equation}

We renormalize the bare interaction strength, $U_{0}$, by requiring
that the two-body $T$-matrices in the quasi-2D and 3D regimes match
when $\lz\rightarrow\infty$, which defines a quasi-2D binding energy
$\bzero$~\cite{Yamashita2014,Petrov2001} as a function of $a_{\text{3D}}$,
with an explicit dependence on $l_{z}$, 
\begin{equation}
B_{0}=\frac{4\hbar^{2}}{ml_{z}^{2}}\arcsinh^{2}\left[\frac{e^{l_{z}/(2a_{\text{3D}})}}{2}\right].\label{eq:b0}
\end{equation}
The binding energy fixes the \emph{BCS-BEC crossover tuning parameter}
$\lza$ which spans from negative (BCS) to positive (BEC) values.
When $\lz\rightarrow0$, the quantity $B_{0}/(2\ef)$ is well defined
and spans the 2D BCS-BEC crossover by introducing 
\begin{equation}
-\text{ln}\left(\kfw\aw\right)=-\text{ln}\sqrt{\frac{B_{0}}{2\ef}}.
\end{equation}

Finally, we can define the density of the system as a function of
two parameters, namely the chemical potential, $\mu$, and the PBC
length, $\lz$, via the number equation 
\begin{equation}
n(\mu,\lz)=-\frac{1}{V}\frac{\partial\Omega(\mu,\lz)}{\partial\mu}\Big|_{\mu;\Delta(\mu)},\label{eq:densityuniform}
\end{equation}
where, for each pair $(\mu,\lz)$, $\Delta(\mu)$ means we have solved
the gap equation before taking the derivative.

\subsection{Local density approximation}

\label{ssec:lda}

As we shall see, the breathing mode frequency in the transverse plane
can be calculated from the integrated 2D density or the so-called
column density, 
\begin{equation}
n_{2D}(\mathbf{\rho=}\sqrt{x^{2}+y^{2}})=\int dz\ n(\rho,z),
\end{equation}
by using a sum-rule approach. We now discuss how to determine the
column density using the uniform density equation of state Eq. (\ref{eq:densityuniform})
and the LDA approach, in the presence of a harmonic trapping potential
in the $xy$-plane and two types of confinement in the axial direction:
\begin{equation}
V_{T}\left(\rho,z\right)=\frac{1}{2}m\omega_{\perp}^{2}\rho^{2}+\left\{ \begin{array}{c}
V_{\infty}\Theta\left[\left|z\right|-l_{z}/2\right],\\
\frac{1}{2}m\omega_{z}^{2}z^{2},
\end{array}\right.
\end{equation}
where the potential $V_{\infty}\Theta[\left|z\right|-l_{z}/2]$, with
$V_{\infty}\rightarrow\infty$ and step function $\Theta(x)$, 
simulates a hard-wall box confinement that may be realized in future
experiments and $m\omega_{z}^{2}z^{2}/2$ is the standard harmonic
trapping potential \cite{Fenech2016,Hueck2018}. In both cases, the
trap aspect ratio, characterized by $\lambda=\hbar/(m\omega_{\bot}l_{z}^2)$
under the hard-wall confinement and $\lambda=\omega_{z}/\omega_{\bot}$
in the case of harmonic potential, should be much larger than $1$.

To calculate the column density, let us first clarify the different
dimensional regimes. In Ref. \cite{Toniolo2017b} the dimensional
crossover of a PBC system, which could be used to describe a nearly
homogeneous quasi-2D Fermi cloud under hard-wall confinement, is split
into three regimes. These are distinguished through the position of
the maximum of the superfluid critical velocity, $v_{c}^{\max{}}$, which has
a non-trivial dependence on the dimensional parameter $\eta$. For
$\eta\leq2$ the maximum of the superfluid velocity is logarithmically
dependent on $\eta$, and we denote this as the 2D regime. The maximum
of the superfluid velocity becomes linear in $\eta$ when $\eta\geq8$
marking the 3D regime. The non-monotonic region contained between
the 2D and 3D regimes is the quasi-2D regime. For further details
on other choices of characterizing the crossover, see Supplemental
Material in Ref. \cite{Toniolo2017b}. 

To understand the dimensional crossover in the presence of a tight
harmonic axial trapping potential, we may determine an equivalent
PBC length scale by approximating
\begin{equation}
l_{z}\sim l_{z}^{\text{HO}}=\sqrt{\frac{\hbar}{m\omega_{z}}}.
\end{equation}
By doing so, Eq. (\ref{eq:densityuniform}) depends on $\lz$ as an
external parameter fixed by the axial harmonic frequency $\omega_{z}$.
We then compare $\lz$ with $\kft$ and obtain a simple relation to
compare the PBC to the harmonically trapped system,
\begin{equation}
\eta\sim\kft l_{z}^{\text{HO}}=\sqrt{\frac{2\eft}{\hbar\omega_{z}}}.\label{eq:yitaHO}
\end{equation}
The single-particle criterion for the harmonically trapped Fermi gas
in the 2D regime is given by requiring: (1) $k_{B}T\ll\epsilon_{F}$
to avoid thermal excitations of the axial harmonic oscillator ground
state, and (2) $\hbar\omega_{z}>\ef$ to ensure that the whole system
is contained in the ground state. By solving $n_{\text{3D}}=n_{f}$,
we see that from Eq. (\ref{eq:densityq2d}) we always have $\ef<\eft$
for $\eta<3\pi/2$. By taking $\kft l_{z}^{\text{HO}}<\sqrt{2}$,
we may interpret $\kft\lz<\sqrt{2}$ as a good approximate regime
of the 2D limit for the trapped case. We denote this regime as the
harmonic oscillator (HO) 2D regime. This distinguishes between the
PBC 2D regime and the 2D regime for a harmonically trapped Fermi gas. 

In Fig. \ref{fig1_bmUnitaryHardWallConfinement} and Fig. \ref{fig2_bmUnitaryHarmonicPotential},
we show the the dimensional regimes as a function of $\eta$ and differentiate
the PBC and harmonically trapped 2D regimes using different colors.
We note that, the harmonically trapped Fermi gas density has been %
experimentally studied \cite{Dyke2011,Dyke2016}. A plateau in the
column density has been observed, by decreasing the total number of
atoms to reach the 2D regime at a given 3D $s$-wave scattering length
\cite{Dyke2016}. By converting the experimentally determined threshold
number density to the dimensional parameter (i.e., using Eq. (\ref{eq:yitaHO})),
we qualitatively determine the boundary of the HO 2D regime as a function
of the interaction strength $(k_{F}^{3D}a_{3D})^{-1}$. This is illustrated
in Fig. \ref{fig3_bmCrossover}(a) {by the pink shaded region}.

\subsubsection{The in-plane LDA}

For a hard-wall confinement along the axial direction, the density
distribution is nearly uniform as a function of $z$ (i.e., see Supplemental
Material in Ref. \cite{Toniolo2017b}). Our theory of a homogeneous
strongly interacting Fermi gas at the dimensional crossover, as outlined
in Sec. \ref{ssec:dimensionalcrossover}, could be quantitatively
applicable. Thus, we must have the column density,
\begin{equation}
n_{2D}\left[\mu\left(\rho\right)\right]=l_{z}n\left[\mu\left(\rho\right),l_{z}\right],\label{eq:n2DHW}
\end{equation}
where $n[\mu(\rho),l_{z}]$ can be calculated using Eq. (\ref{eq:densityuniform}),
once a local chemical potential $\mu(\rho)$ at the radius $\rho$
is provided. For a slowly varying transverse potential $m\omega_{\perp}^{2}\rho^{2}/2$,
the assignment of a local chemical potential is a well-established
approximation, as the surface energy related to the potential change
becomes negligible compared to the bulk energy scale. This treatment
is known as the Thomas-Fermi approximation or LDA. More explicitly, we
have a local chemical potential in Eq. (\ref{eq:n2DHW}):
\begin{equation}
\mu\left(\rho\right)=\mu_{g}-\frac{1}{2}m\omega_{\perp}^{2}\rho^{2},
\end{equation}
where the chemical potential at the trap center $\mu_{g}$ should
be adjusted to yield the total number of atoms $N$, i.e.,
\begin{equation}
N=2\pi\int_{0}^{\infty}\rho n_{2D}\left(\rho\right)d\rho.\label{eq:Ntot}
\end{equation}
In the following, this LDA scheme is referred to as the \emph{in-plane}
LDA.

\subsubsection{The all-direction LDA}

The situation becomes much more complicated for a harmonic axial trapping
potential. This soft-wall potential allows density variation in the
$z$-direction. It is clear that the density distribution of the Fermi
cloud could have very different $z$-dependence at different dimensional
regimes. Deep in the 2D regime, we anticipate that the density profile
may be approximated by,
\begin{equation}
n\left[\mu\left(\rho,z\right)\right]\simeq n_{2D}\left(\rho\right)\left|\Phi_{0}\left(z\right)\right|^{2},
\end{equation}
where $\Phi_{0}\left(z\right)$ is the ground state HO wave-function
along the $z$-direction that is normalized to unity, i.e., $\int dz\left|\Phi(z)\right|^{2}=1$.
As the confinement is tight along the $z$-direction, we are of course
not allowed to define a $z$-dependent local chemical potential and
use Eq. (\ref{eq:densityuniform}) to calculate $n_{2D}\left(\rho\right)$.
However, there is an interesting observation in the deep 2D regime.
As all the atoms are confined in the ground state of the tight confinement,
the in-plane motion of the atoms should be universally described by
the same 2D Hamiltonian, regardless of the detailed form of the confinement.
This implies that the 2D density equation of state $n_{2D}(\mu)$
should be independent of the form of tight confinement, as far as
the confinement gives the same 2D binding energy or 2D scattering
length. Therefore, we could still use Eq. (\ref{eq:n2DHW}) to determine
the column density, provided that the length $l_{z}$ is accurately
approximated in the presence of the axial harmonic trapping potential.

Away from the deep 2D limit, we expect this approximation to increasingly
fail in describing the harmonically confined system when the dimensional
parameter $\eta$ moves towards the quasi-2D and 3D regimes of the
PBC confined model. Fortunately, in the deep 3D regime, the axial
trapping potential $m\omega_{z}^{2}z^{2}/2$ becomes slowly varying
in space as well. In this case, we may implement an \emph{all-direction}
LDA scheme, by setting 
\begin{equation}
\mu(\rho,z)=\mu_{g}-\frac{1}{2}m\omega_{\perp}^{2}\rho^{2}-\frac{1}{2}m\omega_{z}^{2}z^{2}.
\end{equation}
We can introduce a new set of variables, $\xi^{2}=\rho^{2}+\lambda^{2}z^{2}$
and $\tan\psi=\lambda z/\rho$, and rewrite the chemical potential
as a function of $\xi$ only, $\mu(\xi)=\mu_{g}-m\omega_{\perp}^{2}\xi^{2}/2$,
for a fixed aspect ratio $\lambda$. The number of particles, $N$,
of the system approximated with LDA, in cylindrical coordinates, is
given by 
\begin{equation}
N=2\pi\int_{-\infty}^{\infty}dz\int_{0}^{\infty}d\rho\ \rho\ n\left[\mu(\rho,z)\right].\label{eq:density_n}
\end{equation}
The above equation can be used as well for the in-plane LDA to replace
Eq. (\ref{eq:Ntot}), if we require $n(\rho,\lz)=n(\rho)$ when $z\in[-\lz/2,\lz/2]$
and $n(\rho,\lz)=0$ otherwise.

As a brief summary, in the presence of an axial harmonic trapping
potential, we will use the in-plane LDA in the 2D regime and the all-direction
LDA in the 3D regime, as an accurate description for the column density
equation of state. At the 2D-3D crossover, we take interpolation between
these two limits and obtain a \emph{qualitative} description.

\subsection{Polytropic column density equation of state}

\label{ssec:polytropicEoS}

In some limiting cases, the column density may be well approximated
by a polytropic form 
\begin{equation}
\mu(n_{2D})\propto n_{2D}^{\gamma},
\end{equation}
which, as we shall see, provides a significant simplification in understanding
the breathing mode. For example, in the deep 2D limit, the weak violation
of the scale invariance implies that \cite{Taylor2012,Hofmann2012}
\begin{equation}
\gamma_{2D}\sim1,
\end{equation}
regardless of the type of the tight axial confinement. In the 3D regime,
if we consider a unitary Fermi gas, the well-known relation $\mu=\xi\varepsilon_{F}\propto n^{2/3}$,
where $\xi$ is the Bertsch parameter \cite{Bloch2008}, gives rise
to,
\begin{eqnarray}
\gamma_{3D}^{(\textrm{HW})} & = & 2/3,\\
\gamma_{3D}^{(\textrm{HO})} & = & 1/2,
\end{eqnarray}
where the superscripts ``HW'' and ``HO'' distinguish the hard-wall
and harmonic axial trapping potentials. We note that, the polytropic
coefficient with harmonic axial trapping potential decreases to $1/2$,
due to the implementation of the LDA along the $z$-axis \cite{Hu2014,DeRosi2015}.

\subsection{Breathing mode frequency}

\label{ssec:breathingmode}

Once we calculate the density as a function of the chemical potential
and position, the collective oscillations of the Fermi gas can be
derived from the hydrodynamic treatment of the system~\cite{Griffin97,Menotti2002}.
These techniques have been successfully employed to predict a large
variety of collective oscillations for fermionic systems~\cite{Csordas2006,Taylor2008,DeRosi2015}.
In this work, we adopt the commonly-used sum-rule method \cite{Stringari1998,Gao2012},
where the breathing mode frequency, $\omega_{B}$, is given by the
ratio 
\begin{equation}
\hbar^{2}\omega_{B}^{2}=\frac{M_{1}}{M_{-1}}.
\end{equation}
$M_{1}$ is given by the energy weighted moment of the density (second
order central momentum of the density distribution), $M_{1}=2N\hbar^{2}\langle\rho^{2}\rangle/m$,
and $M_{-1}$ is related to a perturbation of the radial coordinate,
$M_{-1}=N\delta\langle\rho^{2}\rangle/\epsilon$, where $\delta\langle\rho^{2}\rangle$
represents the second order momentum when the transverse harmonic
oscillator potential is perturbed by $-\epsilon\rho^{2}$. The expectation
value of the radius squared is given by, 
\begin{equation}
\langle\rho^{2}\rangle\propto\int_{0}^{\infty}\rho^{3}n_{2D}(\rho)d\rho\label{eq:densitycylindrical}
\end{equation}
and we can recast the perturbation of the radial coordinate to a perturbation
of $\omega_{\perp}$, obtaining the closed form~\cite{Menotti2002}
\begin{equation}
\hbar^{2}\omega_{B}^{2}=-2\langle\rho^{2}\rangle\left[\frac{d\langle\rho^{2}\rangle}{d(\omega_{\perp}^{2})}\right]^{-1}.\label{eq:breath}
\end{equation}

From Eq. (\ref{eq:breath}) we observe that we need to know $\langle\rho^{2}\rangle$
up to any constant which doesn't implicitly depend on $\omega_{\perp}$.
We show in Appendix \ref{app:schemes} that the number of particles
falls out of the computation of $\omega_{B}$ when we dimensionalize
the results with the transverse frequency $\omega_{\perp}$. The right-hand-side
of Eq. (\ref{eq:breath}) is expected to be linear in $\omega_{\perp}^{2}$
and return a constant when we evaluate $\omega_{B}/\omega_{\perp}$
(for further details see Appendix \ref{app:schemes}). 

It is worth noting that, when the equation of state has a polytropic
form, $\mu(n_{2D})\propto n_{2D}^{\gamma}$, the sum-rule approach
for evaluating the breathing mode frequency become exact \cite{Hu2014,DeRosi2015}.
It gives (see Appendix \ref{app:schemes} for the derivation),
\begin{equation}
\frac{\omega_{B}}{\omega_{\perp}}=\sqrt{2+2\gamma}.
\end{equation}
Therefore, we anticipate that in different dimensional regimes the
breathing mode frequency may behave like,
\begin{eqnarray}
\omega_{B,2D} & \sim & 2\omega_{\perp},\\
\omega_{B,3D}^{(\textrm{HW})} & = & \sqrt{10/3}\omega_{\perp},\\
\omega_{B,3D}^{(\textrm{HO})} & = & \sqrt{3}\omega_{\perp}.
\end{eqnarray}
The latter two results hold for a unitary Fermi gas only.
\begin{figure}
\centering{}\includegraphics[width=0.48\textwidth]{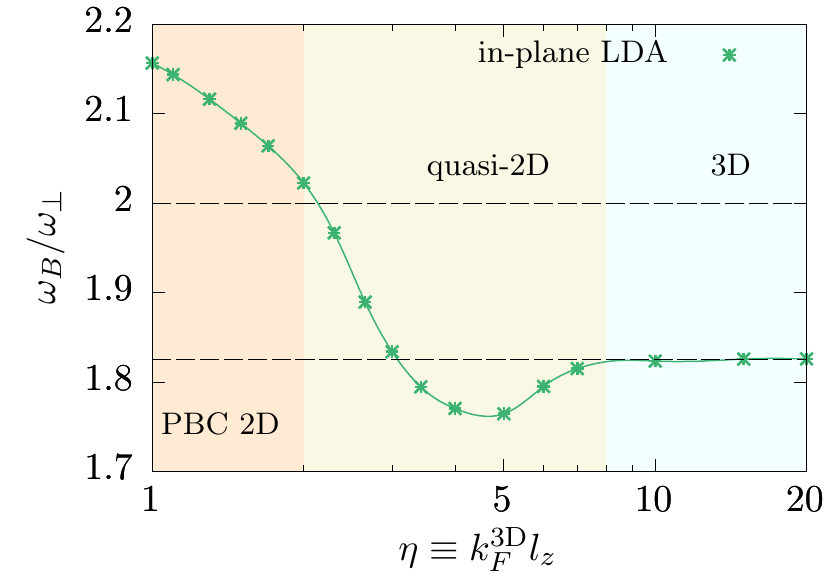}
\caption{\label{fig1_bmUnitaryHardWallConfinement}The breathing mode frequency
$\omega_{B}$ in units of $\omega_{\perp}$ as a function of the dimensional
crossover tuning parameter $\eta\equiv\kft\lz$, when the BCS-BEC
crossover is tuned at unitarity with $\at=\infty$. Here, we consider
the hard-wall confinement along the axial direction. The upper and
bottom dashed lines are the scale invariant predictions in the 2D
and 3D limits, $\omega_{B,2D}=2\omega_{\perp}$ and $\omega_{B,3D}^{(\textrm{HW})}=\sqrt{10/3}\omega_{\perp}\simeq1.83\omega_{\perp}$,
respectively.}
\end{figure}

\section{Results}

\label{sec:results}

We now report the breathing mode frequency at the dimensional crossover
and consider the two different types of axial confinement: the hard-wall
box trapping potential and soft-wall harmonic potential. The former
case is only briefly discussed, as the hard-wall confinement is yet
to be experimentally demonstrated. Hereafter, without any confusions
we use $k_{F}^{3D}\equiv(3\pi^{2}n_{0})^{1/3}$ to represent the 3D
Fermi momentum of an \emph{interacting} Fermi gas at the trap center
with density $n_{0}\equiv n(\rho=0,z=0)$. In our case of considering
two different axial confinements, this turns out to be a more convenient
option than the use of the 3D Fermi momentum of an ideal Fermi gas
at the trap center.

\begin{figure}
\centering{}\includegraphics[width=0.48\textwidth]{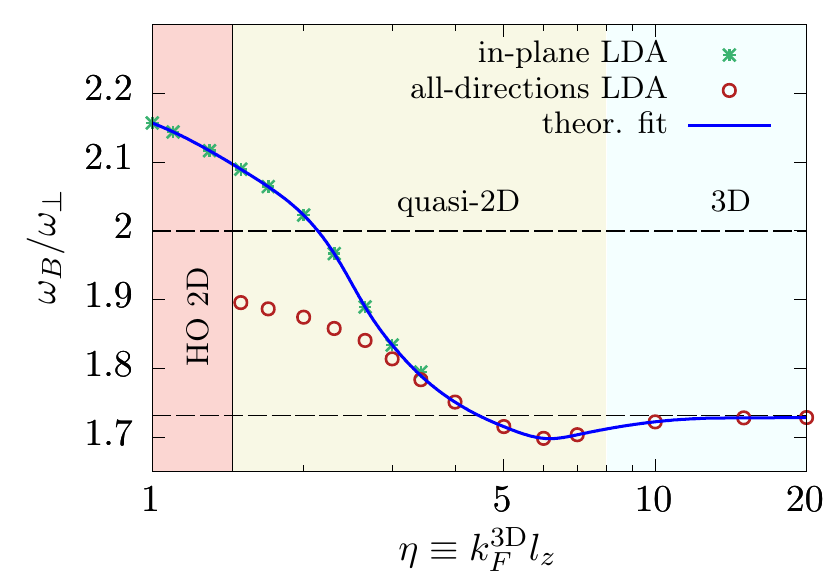}
\caption{\label{fig2_bmUnitaryHarmonicPotential}The breathing mode frequency
$\omega_{B}$ divided by $\omega_{\perp}$ as a function of the dimensional
crossover tuning parameter $\eta\equiv\kft\lz$ when the BCS-BEC crossover
is tuned at unitarity with $\at=\infty$. The results from the in-plane
LDA (green stars) and all-direction LDA (brown circles) schemes are
merged when the lines match to form a qualitative fit (blue solid).
The dimensional crossover is divided into the three dimensional regimes
as in Ref. \cite{Toniolo2017b} and the HO 2D regime is determined
using the experimental criterion of Ref. \cite{Dyke2011} for a 2D
harmonically trapped Fermi gas. The upper and bottom dashed lines
are the scale invariant predictions in the 2D and 3D limits, $\omega_{B,2D}=2\omega_{\perp}$
and $\omega_{B,3D}^{(\textrm{HO})}=\sqrt{3}\omega_{\perp}\simeq1.73\omega_{\perp}$,
respectively.}
\end{figure}

\subsection{The hard-wall axial confinement}

In Fig. \ref{fig1_bmUnitaryHardWallConfinement} we present the breathing
mode frequency of a unitary Fermi gas at the dimensional crossover,
in the presence of a hard-wall axial confinement. The mode frequency
is calculated by using the in-plane LDA, as a function of the dimensional
parameter $\eta$ expanding from the PBC 2D regime when $\eta\leq2$
to the 3D regime when $\eta\geq8$. As our GPF theory provides reliable
equation of state at the dimensional crossover, we anticipate that
our prediction on the breathing mode frequency {is reliable}. 
In the PBC 2D regime, the mode frequency is larger than $2\omega_{\perp}$,
indicating a pronounced quantum anomaly. As we move to the quasi-2D
regime, the frequency decreases rapidly, reaches a minimum at $\eta\sim5$
and finally approaches the 3D limiting value of $\omega_{B,3D}^{(\textrm{HW})}=\sqrt{10/3}\omega_{\perp}$
at $\eta\geq10$.

\subsection{The harmonic axial confinement}

In Fig. \ref{fig2_bmUnitaryHarmonicPotential}, we show the dimensional
crossover of the breathing mode again for the unitary Fermi gas, but
with the harmonic axial trapping potential. Here, the 3D regime is
reached as before when $\eta\geq8$, and the HO 2D regime is realized
when $\eta\leq\sqrt{2}$. As we mentioned earlier, we calculate the
breathing mode frequency using the in-plane LDA scheme near the 2D
regime (green stars) and using the all-direction LDA scheme close
to the 3D regime (brown circles). The in-plane LDA fails to describe
the 3D regime, so we show its prediction at $\eta<4$ only. The all-direction
LDA scheme fails in the 2D regime, since the ground state wave-function
in the axial direction is essentially a Gaussian. As a guide to the
eyes, we combine the two different LDA schemes with the blue solid
line, and this qualitatively describes the breathing mode frequency
in two mutually exclusive regions of the dimensional crossover. By
increasing $\eta$, we find that the mode frequency shows the same
behavior as in the case of the hard-wall confinement: it decreases
quickly away from the 2D regime, exhibits a minimum in the quasi-2D
regime and then saturates to a 3D limiting value, which is $\omega_{B,3D}^{(\textrm{HO})}=\sqrt{3}\omega_{\perp}\simeq1.73\omega_{\perp}$
in the presence of the harmonic trapping potential.

We now turn to describe the behavior of the breathing mode frequency
at the BEC-BCS crossover other than the unitarity limit. For this
purpose, we need to distinguish different interacting regimes and
\emph{clarify} the so-called unitarity regime. In all the previous
discussions, the unitarity regime and an infinite 3D scattering length
are two exchangeable terminologies, both of which can be used without
any confusions in the 3D regime. Away from the 3D limit, however it
seems more intuitive to define the unitarity regime as the regime
where the coherence length of Cooper pairs is comparable to the inter-particle
distance and where the fermionic superfluidity is most robust.

It is worth noting that, fixing a constant 3D interacting parameter
is the best way to compare with experimental results, however from
a theoretical point of view we want an interaction parameter which
probes the same interacting regime as a function of $\eta$. If we
choose the simple condition for the 3D interacting parameter, i.e.
$(\kft\at)^{-1}$ fixed equal to a constant, when we span the dimensional
parameter $\eta$, the system crosses different interacting regimes.
For example, in Fig. \ref{fig1_bmUnitaryHardWallConfinement} where
we have set the 3D scattering length to infinity, we are in the unitarity
regime in the 3D limit while the system enters the BEC regime for the quasi-2D
and 2D regimes. In the HO 2D case, the system is even in the deep
BEC regime \cite{Dyke2011}. From now on, we fix the unitary regime
through the maximum of the critical velocity, as described in our
previous work in determining the dimensional regimes \cite{Toniolo2017b}.
This is a reasonable definition, since the maximum critical velocity
implies the most robust fermionic superfluidity.

\begin{figure}
\centering{}\includegraphics[width=0.48\textwidth]{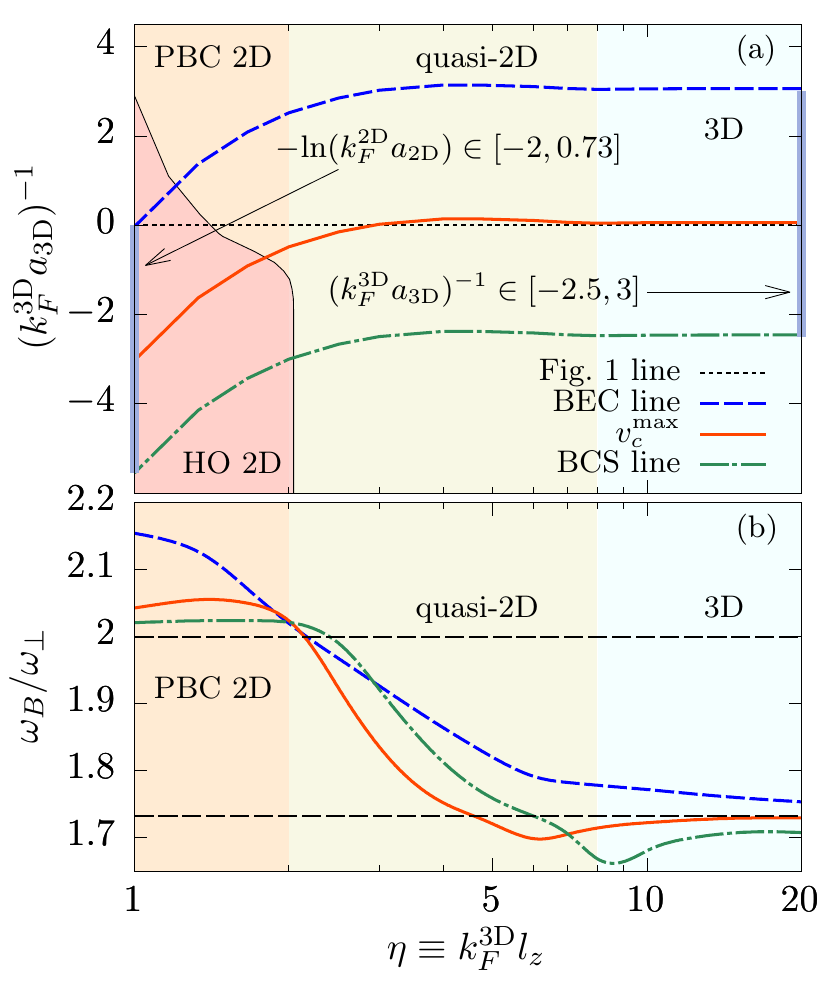}
\caption{\label{fig3_bmCrossover} (a) The interaction parameter $(\kft\at)^{-1}$,
as a function of $\eta$, fixed by the BCS, BEC, and $v_{c}^{\max{}}$
lines in order to span the dimensional crossover and maintain the
system in the BCS (dashed dotted green), unitarity (solid red), and
BEC (dashed blue) interacting regimes. According to Ref. \cite{Toniolo2017b}
the solid red line is taken to be a good criterion to distinguish
the BCS and BEC regimes. (b) we show the breathing mode frequency
$\omega_{B}$ in units of $\omega_{\perp}$ as a function of the dimensional
crossover parameter $\eta\equiv\kft\lz$ for the BCS (dashed dotted
green), unitarity (solid red), and BEC (dashed blue) interacting regimes. }
\end{figure}

Figure \ref{fig3_bmCrossover}(a) displays the choices made for the
BCS (dashed-dotted green line) and BEC (dashed blue line) crossover
regimes, in which the two lines are obtained by vertically shifting
the maximum critical velocity curve down and up by some amounts. These
choices appear to be optimal since they both span the 2D and 3D limits
($\eta=2$ and $\eta=20$ respectively). For the 2D limit both the
PBC 2D regime and the HO 2D regime are reached, and converting the
scattering length to its 2D counterpart, $\ln(\kfw\aw)$, we observe
that it spans the relevant part of BCS-BEC crossover.

In Fig. \ref{fig3_bmCrossover}(b) we plot the ratio $\omega_{B}/\omega_{\perp}$
in the different interacting regimes as per Fig. \ref{fig3_bmCrossover}(a),
the BCS (dashed dotted green), unitarity (solid red), and BEC (dashed
blue) regimes. We see that the deviation of the breathing mode from
the classic result, $\omega_{B}=2\omega_{\perp}$, appears when the
2D region is entered. Since the quantum anomaly is due to the presence
of the renormalization energy $B_{0}$, which tends to vanish while
approaching the BCS regime, we observe a strong deviation in the BEC
regime (dashed blue) which is progressively reduced in the unitarity
regime (solid red). Qualitatively, the fit between the in-plane and
all-direction LDA results drop from $2\omega_{\perp}$ to a range
of values around the 3D unitarity limit, $\omega_{B}=\sqrt{3}\omega_{\perp}$.
The unitarity results (red-solid) converge to this value, while as
remarked in Ref. \cite{Hu2004}, the BEC regime provides a larger
value of $\omega_{B}$, and in the BCS limit there is a non trivial
behavior below $\omega_{B}=\sqrt{3}\omega_{\perp}$.

\subsection{Quantum anomaly in the deep 2D regime}

Focusing on the 2D regime, i.e. $\eta=1$, we compare our results
with previous two-dimensional studies \cite{Hofmann2012,Gao2012}.
Since the choice $\eta=1$ and a large range of values of $\ln(\kfw\aw)$
are contained both in the PBC 2D and harmonic oscillator 2D regime,
we compare the anomaly through the quantity $\delta\omega_{B}/\omega_{\perp}$,
where $\delta\omega_{B}=\omega_{B}-2\omega_{\perp}$. 
\begin{figure}[t!]
\centering{}\includegraphics[width=0.48\textwidth]{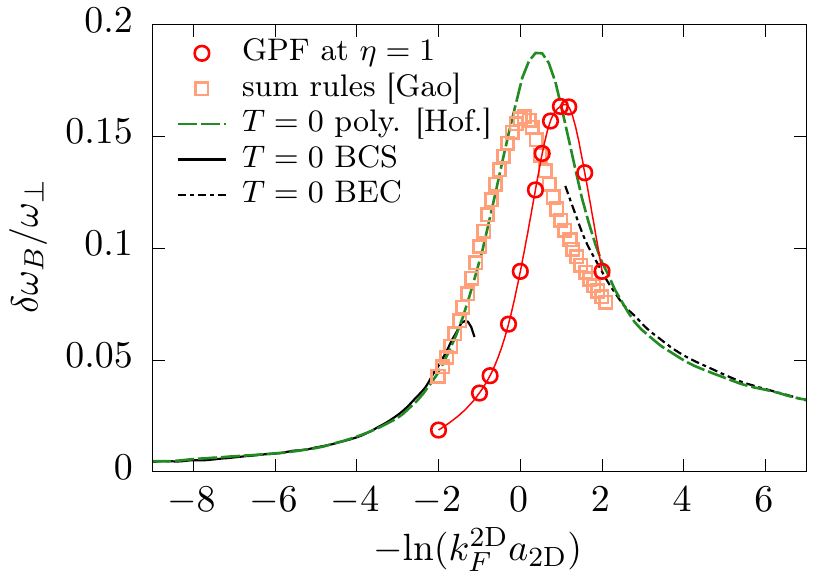} \caption{\label{fig4_bm2D}The quantum anomaly deviation $\delta\omega_{B}=\omega_{B}-2\omega_{\perp}$
of the breathing mode frequency $\omega_{B}$ in units of the transverse
harmonic trapping frequency $\omega_{\perp}$. Our results (red circles)
are obtained by applying the in-plane LDA for the PBC confined 2D
Fermi gas ($\eta=1$) at the GPF level. For comparison, we show also
the polytropic fit from Ref. \cite{Hofmann2012} (dashed green), the
sum rule results from Ref. \cite{Gao2012} (orange squares), and the
zero temperature analytic predictions for the far BCS (solid black)
and BEC (dashed-dotted black) regimes.}
\end{figure}

We observe that the qualitative behavior of the quantum anomaly is
recovered by our data, and the maximum of the deviation, $\delta\omega_{B}$,
is approximately the same height of Ref. \cite{Gao2012}. The shift
of the anomaly to the BEC side in our results is due to either the
GPF contribution to the global chemical potential $\mu_{g}$ in comparison
to the quantum Monte Carlo schemes, or that for $\eta=1$ the range
of $\ln(\kfw\aw)$ is shifted with respect to the exact 2D case when
we consider the exact 2D limit.

\section{Conclusions}

In conclusions, we have characterized the breathing mode of a strongly
interacting Fermi gas at the dimensional crossover from two- to three-dimensions,
as a function of the interatomic interaction strength. Using two schemes
for the local density approximation, through the hydrodynamic formalism
and sum rules we are able to calculate the breathing mode within a
beyond mean-field, gaussian pair fluctuation theory. Two kinds of
tight axial confinements have been considered: a hard-wall box potential
and a soft-wall harmonic trapping potential. In both cases, we have
shown that the quantum anomaly will be visible in the breathing mode
frequency as we approach two dimensions in the strongly interacting
regime. We have compared our breathing mode anomaly in two-dimensions
directly to the previous predictions based quantum Monte Carlo simulations
and have found a good agreement. As the dimension of the system changes
to quasi-2D, the breathing mode decreases in a non-monotonic way,
and towards the 3D regime, it saturates to the anticipated scaling
invariant values, in the case of an infinite three-dimensional scattering
length.

Our results may be quantitatively applicable to the case of the hard-wall
axial confinement, where the density distribution along the axial
direction is more or less uniform. For the case of an axial harmonic
trapping potential, we instead anticipate that our results provide
a good qualitative description, due to ambiguity in interpreting the
length of axial confinement $l_{z}$. We are now working on the density
equation of state by explicitly including harmonic trapping in the
axial direction, and aim to provide a more quantitative description.
\begin{acknowledgments}
We thank Paul Dyke, Sascha Hoinka and Chris Vale for stimulating discussions.
This research was supported by Australian Research Council's (ARC)
Discovery Projects: FT140100003 and DP180102018 (XJL), and FT130100815
and DP170104008 (HH).
\end{acknowledgments}

\appendix

\section{Sum rule for the breathing mode frequency}

\label{app:schemes} 

According to Eq. (\ref{eq:breath}), we need to know $\langle\rho^{2}\rangle$
up to any constant which is not dependent on $\omega_{\perp}$,
this comes from the fact we need to divide {the function} by its own derivative. We
notice also that we must follow different approaches for the in-plane
and the all-direction LDA schemes.

\subsection{The in-plane LDA}

In the in-plane LDA we start from Eq.~(\ref{eq:densitycylindrical})
and impose $n_{2D}(\rho)=l_{z}n(\rho)$, since $n(\rho,z)=n(\rho)$
when $z\in[-\lz/2,\lz/2]$ and vanishing otherwise. We then apply
the LDA and we require $n(\rho)\equiv n[\mu(\rho)]$ where 
\begin{equation}
\mu(\rho)=\mu_{g}-\frac{1}{2}m\omega_{\perp}^{2}\rho^{2}
\end{equation}
and $\mu_{g}$ is a constant. Thus we can compute Eq.~(\ref{eq:density_n})
and Eq.~(\ref{eq:densitycylindrical}) employing a change of variables
from $\rho$ to $\mu$, 
\begin{equation}
-\frac{d\mu}{m\omega_{\perp}^{2}}=\rho d\rho,\qquad\rho=\frac{1}{\omega_{\perp}}\sqrt{\frac{2}{m}(\mu_{g}-\mu)},\label{eq:changeofvariable}
\end{equation}
with $\mu(\rho=0)=\mu_{g}$ and $\mu(\rho=\infty)=-\infty$. We thus
obtain, from Eq.~(\ref{eq:density_n}) 
\begin{equation}
\frac{Nm}{2\pi\lz}\omega_{\perp}^{2}=\int_{-\infty}^{\mu_{g}}d\mu\ n(\mu)\label{eq:newdens}
\end{equation}
which is always convergent, as when $\mu\leq-B_{0}/2$ we have $n=0$.
Also to simplify the notation we introduce the constant $\kappa_{p}=Nm/(2\pi\lz)$
and a new variable $y=\kappa_{p}\omega_{\perp}^{2}$. From the definition
of the density $n(\mu)=-\partial_{\mu}\Omega$, we integrate to obtain,
\begin{equation}
y=-\Omega(\mu_{g}).\label{eq:y1}
\end{equation}
From Eq.~(\ref{eq:newdens}) then we can numerically compute the
dependency of $\mu_{g}$ on $\omega_{\perp}$, via the function $\mu_{g}\equiv\mu_{g}(y)$.
Also by applying the same change of variable as before, we obtain
from Eq. (\ref{eq:densitycylindrical}), 
\begin{equation}
\langle\rho^{2}\rangle\propto-\frac{1}{y^{2}}\int_{-\infty}^{\mu_{g}(y)}d\mu\ \Omega(\mu).\label{eq:rou2}
\end{equation}
Since we have $d/d(\omega_{\perp}^{2})\propto d/dy$, we obtain 
\begin{equation}
\frac{d}{dy}\langle\rho^{2}\rangle\propto-\frac{2}{y}\langle\rho^{2}\rangle-\frac{1}{y^{2}}\frac{d}{dy}\int_{-\infty}^{\mu_{g}(y)}d\mu\ \Omega(\mu).
\end{equation}
We use the fact that $\Omega$ turns out to be always strictly decreasing
monotonically, which means that $y$ is a strictly increasing monotonic
function of $\mu_{g}$ and we can apply the inverse derivative theorem
globally, i.e. 
\begin{equation}
\frac{d}{dy}=\left(\left.\frac{dy}{d\mu_{g}}\right|_{\mu_{g}(y)}\right)^{-1}\frac{d}{d\mu_{g}}\Big|_{\mu_{g}(y)},\label{eq:inversedev}
\end{equation}
which gives 
\begin{equation}
\frac{d}{dy}\langle\rho^{2}\rangle\propto-\frac{2}{y}\langle\rho^{2}\rangle+\frac{1}{y}\frac{1}{n[\mu(y)]}.
\end{equation}
Finally, due to the proportionality constant $\kappa_{p}$, we can
compute 
\begin{equation}
\frac{\omega_{B}^{2}}{\omega_{\perp}^{2}}=-\frac{2}{y}\frac{\langle\rho^{2}\rangle}{d\langle\rho^{2}\rangle/dy}=\left(1-\frac{1}{2n[\mu_{g}(y)]\langle\rho^{2}\rangle}\right)^{-1}.\label{eq:breathingpartial}
\end{equation}

For a polytropic density equation of state, which takes the following
form with the step function $\Theta(x),$
\begin{equation}
n\left(\mu\right)\propto\left(\mu+\frac{B_{0}}{2}\right)^{1/\gamma}\Theta\left[\mu+\frac{B_{0}}{2}\right],
\end{equation}
it is easy to see that 
\begin{eqnarray}
\Omega\left(\mu\right) & \propto & \left(\mu+\frac{B_{0}}{2}\right)^{\left(1+\gamma\right)/\gamma},\\
y & \propto & \left(\mu_{g}+\frac{B_{0}}{2}\right)^{\left(1+\gamma\right)/\gamma},\\
\left\langle \rho^{2}\right\rangle  & \propto & \left(\mu_{g}+\frac{B_{0}}{2}\right)^{\left(1+2\gamma\right)/\gamma},
\end{eqnarray}
by using Eqs. (\ref{eq:y1}) and (\ref{eq:rou2}), respectively. Thus,
we obtain 
\begin{equation}
\left\langle \rho^{2}\right\rangle \propto y^{-1/(1+\gamma)}.
\end{equation}
Using Eq. (\ref{eq:breathingpartial}) we also arrive at the well-known
sum-rule relation,
\begin{equation}
\frac{\omega_{B}^{2}}{\omega_{\perp}^{2}}=2+2\gamma.
\end{equation}

In actual computations, the density equation of state generally does
not follow the idealized polytropic form. Using the GPF theory as
outlined in Sec. \ref{ssec:dimensionalcrossover}, we calculate the
thermodynamic function $\Omega(\mu)$ for a broad range of values
at a given set of parameters (such as $l_{z}$ and 3D scattering length
$a_{3D}$), starting from the minimum chemical potential $-B_{0}/2$
where $\Omega=0$. We then compute the quantity $y^{2}n[\mu_{g}(y)]\langle\rho^{2}\rangle$,
which is quadratic in $y$ by solving Eq. (\ref{eq:y1}). We fit the
results with a quadratic function and extract the second order Taylor
coefficient at each $y$, using this coefficient in Eq. (\ref{eq:breathingpartial})
to directly obtain $\omega_{B}/\omega_{\perp}$. We finally convert
the peak density at the trap center $n_{0}=n[\mu_{g}(y)]$ at the
given $y$ to the 3D Fermi momentum at the trap center $k_{F}^{3D}=(3\pi^{2}n_{0})^{1/3}$
and show the breathing mode frequency $\omega_{B}/\omega_{\perp}$
as a function of the dimensional parameter $k_{F}^{3D}l_{z}$.

\subsection{The all-direction LDA}

The all-direction LDA mirrors the procedures for the in-plane LDA
case. Again we start from the number equation and exploit the symmetry
in $z$, 
\begin{equation}
N=4\pi\int_{0}^{\infty}dz\int_{0}^{\infty}d\rho\ \rho\ n\left[\mu\left(\rho,z\right)\right],
\end{equation}
where we have assumed 
\begin{equation}
\mu(\rho,z)=\mu_{g}-\frac{1}{2}m\omega_{\perp}^{2}(\rho^{2}+\lambda^{2}z^{2}).
\end{equation}
The variables $z$ and $\rho$ span the first quadrant of $\RR^{2}$
and such a surface can be mapped by the polar coordinates $\xi\in[0,\infty]$
and $\psi\in[0,\pi/2]$, defined as 
\begin{equation}
\xi^{2}=\rho^{2}+\lambda^{2}z^{2}\qquad\tan\psi=\frac{\lambda z}{\rho}.
\end{equation}
The number of particles is, 
\begin{equation}
\frac{N\lambda}{4\pi}=\int_{0}^{\infty}d\xi\left[\xi^{2}n(\xi)\right],
\end{equation}
a change of variables, identical to Eq. (\ref{eq:changeofvariable}),
allows us to obtain, 
\begin{equation}
y=\kappa_{c}\omega_{\perp}^{2}=-\int_{-\infty}^{\mu_{g}}d\mu\sqrt{\mu_{g}-\mu}\ \frac{d\Omega}{d\mu},\label{eq:densitycomplete}
\end{equation}
with 
\begin{equation}
\kappa_{c}=\frac{N\omega_{z}}{2\pi}\left(\frac{m}{2}\right)^{3/2}\omega_{z}=\kappa_{p}\frac{\lz\omega_{z}}{2}\sqrt{\frac{m}{2}}.
\end{equation}
By applying $\lz\simeq\sqrt{\hbar/(m\omega_{z})}$, we obtain the
ratio $\kappa_{c}/\kappa_{p}=\hbar/(2\sqrt{2m}\lz)$. With a very
similar procedure we also compute 
\begin{equation}
\langle\rho^{2}\rangle\propto-\frac{1}{y^{2}}\int_{-\infty}^{\mu_{g}(y)}d\mu\ \sqrt{\mu_{g}(y)-\mu}\ \Omega(\mu),
\end{equation}
and then its derivative, 
\begin{equation}
\frac{d\langle\rho^{2}\rangle}{dy}\propto-2y^{-1}\langle\rho^{2}\rangle-y^{-2}\frac{d}{dy}\int_{-\infty}^{\mu_{g}(y)}d\mu\sqrt{\mu_{g}(y)-\mu}\ \Omega(\mu).
\end{equation}
Since $\mu_{g}(y)$ is a monotonic function of $y$ we can invert
the derivative globally by using Eq. (\ref{eq:inversedev}), which
introduces the quantity 
\begin{equation}
I(\mu_{g})=\frac{dy}{d\mu_{g}}=-\int_{-\infty}^{\mu_{g}}d\mu\sqrt{\mu_{g}-\mu}\ \frac{d^{2}\Omega}{d\mu^{2}},
\end{equation}
and, as observed before, 
\begin{equation}
\frac{d}{d\mu_{g}}\int_{-\infty}^{\mu_{g}(y)}d\mu\sqrt{\mu_{g}-\mu}\ \Omega(\mu)=-y.
\end{equation}
Similarly to the in-plane LDA case, we have 
\begin{equation}
\frac{\omega_{B}^{2}}{\omega_{\perp}^{2}}=\left(1-\frac{1}{2I(\mu_{g}(y))\langle\rho^{2}\rangle}\right)^{-1}.\label{eq:freqcomplete}
\end{equation}

The computation of the breathing mode frequency in the all-direction
LDA requires a further step. We are going to fit quadratically the
function $y\mapsto y^{2}I(\mu_{g}(y))\langle\rho^{2}\rangle$ and
obtain the second order Taylor coefficient, as for the in-plane LDA, but
we need to consider an important subtlety, the number of particles,
$N$, was hidden by the $y$ variable in both in-plane and all-direction
schemes and these need to be the same in order to make a consistent
comparison in the case of the harmonic axial trapping potential. Eq.
(\ref{eq:densitycomplete}) needs to be computed for a fixed $\kappa_{p}$
and then the ratio $\kappa_{c}/\kappa_{p}$ adjusted according to
the choice of $\lz$ we are considering. By doing so we are not modifying
the form of Eq. (\ref{eq:freqcomplete}), but only adjusting $\mu_{g}$
to the correct number of particles (which is never explicit but fixed)
in Eq. (\ref{eq:densitycomplete}).

\bibliographystyle{apsrev4-1.bst}
\phantomsection\addcontentsline{toc}{section}{\refname}\bibliography{main_bib_2}
 
\end{document}